\documentstyle[twocolumn, aps,psfig]{revtex}

\newcommand{\be}{\begin{equation}}
\newcommand{\ee}{\end{equation}}
\newcommand{\bea}{\begin{eqnarray}}
\newcommand{\ba}{\begin{array}}
\newcommand{\eea}{\end{eqnarray}}
\newcommand{\ea}{\end{array}}

\newcommand{\s}{\sigma}

\begin{document}

\twocolumn[\hsize\textwidth\columnwidth\hsize\csname@twocolumnfalse%
\endcsname

\title{Criticality and oscillatory behavior in non-Markovian Contact Process}
                              
\author{Rouzbeh Gerami}  

\address{Institute for Studies in Theoretical Physics and Mathematics
,PO Box 19395-5531, Tehran, Iran.}

\address{ Department of Physics, Sharif University of Technology
,PO Box 11364-9161, Tehran, Iran.}

\address{ Department of Physics and Astronomy, University of 
Southern California, Los Angeles, CA 90089-0484, USA.}

\date{\today}

\maketitle

\begin{abstract}
A Non-Markovian generalization of one-dimensional Contact Process (CP) is being 
introduced in which every particle has an age and will be annihilated at its 
maximum age $\tau$. There is an absorbing state phase transition which is 
controlled by this parameter. The model can demonstrate oscillatory behavior 
in its approach to the stationary state. These oscillations are also present
in the mean-field approximation which is a first-order differential equation
with time-delay. Studying dynamical critical exponents suggests that the model
belongs to the DP universlity class.

\end{abstract}

]

%%%%%%%%%%%%%%%%%%%%%%%%%%%%%%%%%%%%%%%%%%%%%%%%%%%%%%%%%%%%%%%%%%%%%%%%

\section{introduction}

Studying phase transitions in the systems far from equilibrium
has been a topic of growing interest in recent years
\cite{Hinrichsen00a,Marro99}. Specially systems with absorbing states
which can not evolve further once they are trapped in one such state
have been an interesting subject of research. Various models with one
or more absorbing states have been studied which belong to a few
universality classes and mostly to that of Directed Percolation (DP).
According to the DP conjecture \cite {Grass82,Janssen81}, every phase
transition in a system with a single absorbing state having short range
interactions with no special symmetry or quenched disorder
\cite{Moreira96} belongs to the DP class. There has been examples of
absorbing state phase transitions without some of the DP conjecture
conditions that still belong to the DP class such as systems with an
infinite number of absorbing states \cite{Jensen93a,Jensen93b}. There
has also been another universality classes for parity conserving systems
\cite{Hinrichsen97a} and as recently proposed, for systems with
infinitely many absorbing states coupled to a conserved field
\cite{Rossi00,Pastor00}.

Although the Markovian property have been implicitly accepted in these
models it is not essential for a nonequilibrium phase transition.
Usually, adding some kind of memory to the system, such that the system
should refer to its history in order to define its future, gives rise to
some new interesting behaviors that are absent in Markovian ones 
\cite{Ohira00,Gerami}. However properties of non-Markovian nonequilibrium 
phase transition have not been studied.

In this paper a non-Markovian variant of the Contact Process (CP)
\cite{Harris74} is introduced and the critical behavior is
investigated. Standard CP in its continuous time version is a
lattice model in which every empty site is being occupied by a
particle with rate  $\lambda n / z$ and every particle is removed
with rate one, where $z$ is the coordination number, $n$ number of
occupied nearest neighbors and $\lambda$ a positive parameter
controlling the creation rate. The system has a second-order
critical point at $\lambda_{c} =3.2978$ and it belongs to the DP
class \cite{Marro99}.

In this model I introduce a {\em memory} for each particle: every particle
knows when it has been created. Like standard CP, every site is being
occupied by a particle at a rate proportional to the number of its occupied
nearest neighbors, while every existing particle will die exactly at age
$\tau$. For large values of $\tau$ the particles live enough to reproduce plenty
of another ones and the system can remain in its active state. As $\tau$ is
decreased, each particle has less time to create another particles and
for $\tau < \tau_{c}$ the system will be trapped in its absorbing state with
probability 1, where there is no existing particle and no new particles
can be born.

Attributing age to the particles in the CP model has been suggested
earlier \cite{Dorogovtsev00}, but not in a way that leads to a non-Markovian
model. Although some interesting alterations in the dynamical behavior
 of the system have been observed.

Non-Markovian property gives rise to some oscillations in the density of
particles. These oscillations are also supported by the Mean-Filed 
approximation. Because of the non-Markovian property of the
system, there is a delay parameter in the mean field equation and
this reproduces the oscillatory solutions observed in the simulations.
By mean-field approach, existence of the phase transition is
justified and a critical age can be found, but like standard CP the critical 
behavior is not described completely.

In this paper, the critical behavior of the model is studied using the
time-dependent Monte Carlo method. The critical dynamical exponents has
been calculated, and shown to be in good agreement of those of DP.

\section{Model}

The model is defined on a 1-dimensional lattice and with continuous time.
Every site is either empty or occupied by a single particle. There is a
chance for a vacant site to be occupied provided there are occupied sites
in its nearest neighborhood. A new particle is born in an empty site with
rate $n/2$ ($n$ number of occupied nearest neighbors). Every particle will
die exactly at time $\tau$ after its birth.

Obviously there must be a phase transition in the system with density of the
particles as the order parameter. For small values of $\tau$ particles
die fast and eventually the system is trapped in its absorbing state. For
large $\tau$'s, particles have a large lifetime and reproduce sufficiently
other ones to keep the system active. Fig. 1 shows a single cluster in
a realization of the model for $\tau=3.5$ and up to $t=100$. 
As can be seen, every particle has a definite lifetime and here the 
system is in its active state.
\begin{figure}
\psfig{figure=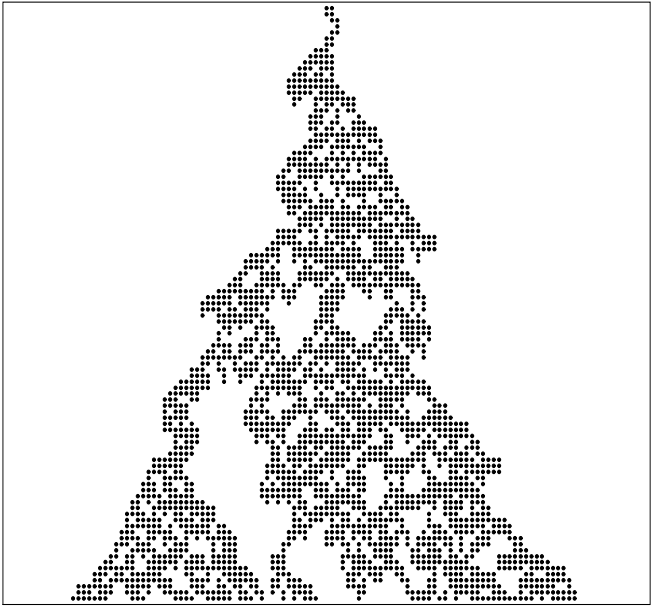,width=8.6cm,height=8cm}
\vspace{0.3cm}

\caption{A typical space-time cluster started with a single
particle in the origin for $\tau = 3.5$ and up to $t=100$. }
\end{figure}

Like CP, here the creation process is not history-dependent, however the
death process is, and thus we have a non-Markovian process. To find out that
how many particles are removed at each moment, we should know how many
particles had been created at time $\tau$ earlier. Therefore a knowledge of
only the present state of the system is not enough to find out the future
states.

\section {Mean-Field equation}

Let $\s_{1}(x)$ and $\s_{0}(x)$ denote the state of site $x$, and $\rho$
be the density of particles. $\s_{1}(x)$ is 1 if the site is occupied and
0 if it is vacant, and we have
\be
\rho = \langle \s_{1} (x) \rangle _{x}  =  1 - \langle \s_{0} (x) \rangle _{x}.
\ee
Therefore the rate of reproduction is
\be
 \langle \s_{1}(x) \s_{0}(x+1) \rangle _{x}
\ee
it can be written in terms of the density-vacancy correlation function (the 
correlation of the of occupied and vacant sites)  
\be
C_{10}(\delta) = \frac {\langle \s_{1}(x) \s_{0}(x+\delta) \rangle _{x}  - \rho(1-\rho)}
  {\rho(1-\rho)}
\ee
Hence the rate of reproduction is 
\be 
 r \rho_{t}(1-\rho_{t}) 
\ee 
where 
\be 
r = 1 + C_{10}(\delta=1). 
\ee

Obviously $C_{10}(\delta=1)$ is negative (a particle reduces the chance of its
nearest neighbors to be vacant), and thus $r<1$. In the mean-field
approximation the correlation is neglected and we put $r=1$.
Therefore the mean-field equation will be
\be
\frac{d\rho_{t}}{dt} =  \rho_{t}(1-\rho_{t}) -
                        \rho_{t-\tau}(1-\rho_{t-\tau}) \hspace{0.7cm} (t>\tau)
\ee
where the second term is the rate of annihilation at time $t$, equal to the
rate of creation at time $t-\tau$. This equation is true for $t>\tau$.
I assume that all existing particles at $t=0$,
gradually die during the time interval $(0,\tau)$. So for $t<\tau$ we have:
\be
\frac{d\rho_{t}}{dt} =  \rho_{t}(1-\rho_{t}) - \rho_{0}/\tau \hspace{1cm} (t \leq \tau).
\ee

This equation can be re-written in the integral form. First by integrating
Eq. (7)
\be
\rho_{t} =  \int_{0}^{t} \rho_{t'} (1-\rho_{t'}) dt'  -  \rho_{0}t/\tau + \rho_{0}
\hspace{0.7cm} (t\leq \tau)
\ee
specially for $t=\tau$
\be
\rho_{t=\tau} =  \int_{0}^{\tau} \rho_{t'} (1-\rho_{t'}) dt'.
\ee

By integrating eq. (6) and making use of eq. (9) we find
\be
\rho_{t} =  \int_{t-\tau}^{t} \rho_{t'} (1-\rho_{t'}) dt' \hspace{1cm} (t > \tau)
\ee
It is a definite integral with a time dependent lower- and upper-limit. 
So although the integrand is non-negative, $\rho(t)$ may have a 
non-monotonic behavior.

Finding stationary density is not possible in the differential equation.
Setting $d\rho_{t}/dt = 0 $ leads to nothing more than $\rho_{t} = 1
- \rho_{t-\tau}$ or $\rho_{t} =\rho_{t-\tau}$. The former is
irrelevant in the steady state and the latter is correct for
every value of $\bar{\rho}$. However by setting 
$\rho_{t} = \rho_{t'} = \bar{\rho}$ in the integral equation (Eq. (10) ), 
it turns out that 
\be
\bar{\rho} = 0   
\ee
or 
\be 
\bar{\rho} = 1 - 1 /  \tau.
\ee 
Thus there is a phase transition at $\tau = \tau_{c} = 1 $.

\section{Oscillations}

It is being observed that the density of particles $\rho$ undergoes 
damped oscillations while approaching the stationary state. The solid 
line in the figure 2 shows one such oscillatory evolution of $\rho$. 
Simulations are done in a lattice of 10000 sites with periodic boundary 
condition. The plotted curves are averaged over 100 realizations for 
$\tau = 7$. Period of oscillations is slightly greater than $\tau$.
This kind of oscillations are present for all values of $\tau$, 
but they are weaker for smaller $\tau$.

\begin{figure}
\psfig{figure=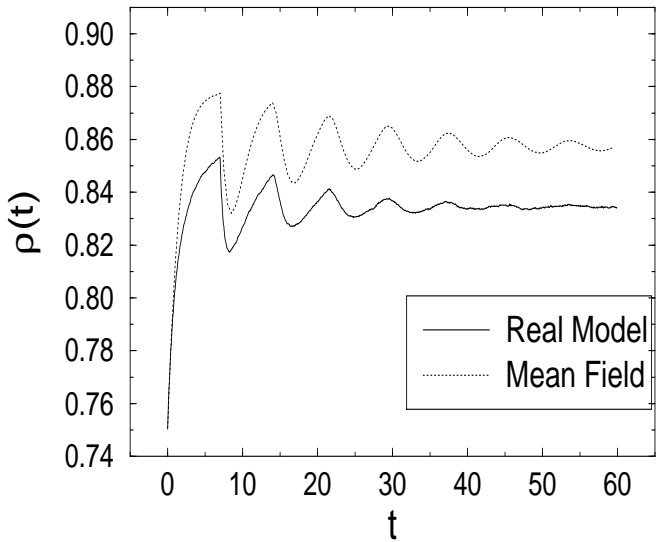,width=8.6cm,height=7.5cm}
\caption{ Density of particles versus time for Non-Markovian
CP in real model (solid line) and in Mean-Field approximation
(dashed line) for $\tau= 7$ and $\rho_{0} = 0.75$. }
\end{figure}

The oscillatory behavior can be understood by paying attention to the 
history-dependence feature of the model. Since every particle dies at 
age $\tau$, the time evolution at time $t$ is coupled to the state of the 
system  at time $t-\tau$. A high creation rate at time $t$ is equivalent to a high 
annihilation rate at time $t+\tau$. So an increase in the density at 
time $t$ can lead to a decrease in it at some time later and naturally 
the period of the oscillations is of order $\tau$.

These oscillations are also supported by the mean-field approximation.
The mean-field equation is a delayed nonlinear first-order differential
equation that is able to have oscillatory solutions which do not exist 
in ordinary first-order differential equations. Figure 2 (dashed line)
shows one of these oscillatory solutions for $\tau=7$. Period of oscillations 
in real model and Mean-Field are the same.

The delay time in the mean-field differential equation, can be eliminated by
a Taylor expansion of $\rho_{t-\tau}$. It basically contains derivatives up to
infinite order. In fact here we have an infinite-order differential equation
which is naturally able to demonstrate many complex behaviors. As in simulations,
Oscillatory behavior is sensitive to changes in the value of $\tau$. It disappears
for small enough values of $\tau$ and it is less damped for larger $\tau$'s.

\section {Critical Behavior}

In this section I present the results concerning the critical behavior of 
the system obtained from simulating the model. Simulation is made using the 
time-dependent Monte Carlo method \cite{Grass79}. In this method, the simulation 
is started with the system in a state very close to the absorbing state that 
is all the sites are vacant except one in origin which is occupied. The age 
of this single particle is initially set to 0. 
The sites 
are updated parallel and after every time increment all existing particles become 
older by that amount. They die after growing up to age $\tau$. 

I measure the average population of particles $N(t)$, averaged over all realizations, 
$P(t)$, the probability of not entering the absorbing state up to time $t$, and 
$R^{2}(t)$ the mean square spreading distance. As a result of the scaling hypothesis
\cite{Grass79}, at criticality, these quantities should scale algebraically as
\be
N(t) \sim t^{\eta}
\ee
\be
P(t) \sim t^{-\delta}
\ee
\be
R^{2}(t) \sim t^{z}
\ee
So at criticality the log-log plot of these functions should asymptotically become
a straight line with the slope equal to the dynamical critical exponents. The local 
slopes for the survival probability $P(t)$ are defined by
\be
\delta (t) = - \frac {\ln[P(t)/P(t/b)]}{\ln(b)}
\ee
and similarly for the other exponents. I usually use $b=15$.
Away from criticality there are either upward or downward curvatures in the 
log-log plot of functions versus $t$ and also in the plot of the critical exponents 
versus $1/t$, depending upon the super- or sub-criticality of the system. By 
detecting the straight line from the curved ones, the value of $\tau_{c}$ can be 
evaluated with a good precision. Having $\tau_{c}$, the critical exponents 
can also be found. 

Simulations are typically done up to time 400 (although many runs enter the absorbing 
state earlier) with a time increment of 0.004 for continuous-time simulation. 
Obviously this is equal to the maximum precision possible
in determining $\tau_{c}$. Statistical quantities have been generally averaged over
10000 different realizations of the model. 

\begin{figure}
\psfig{figure=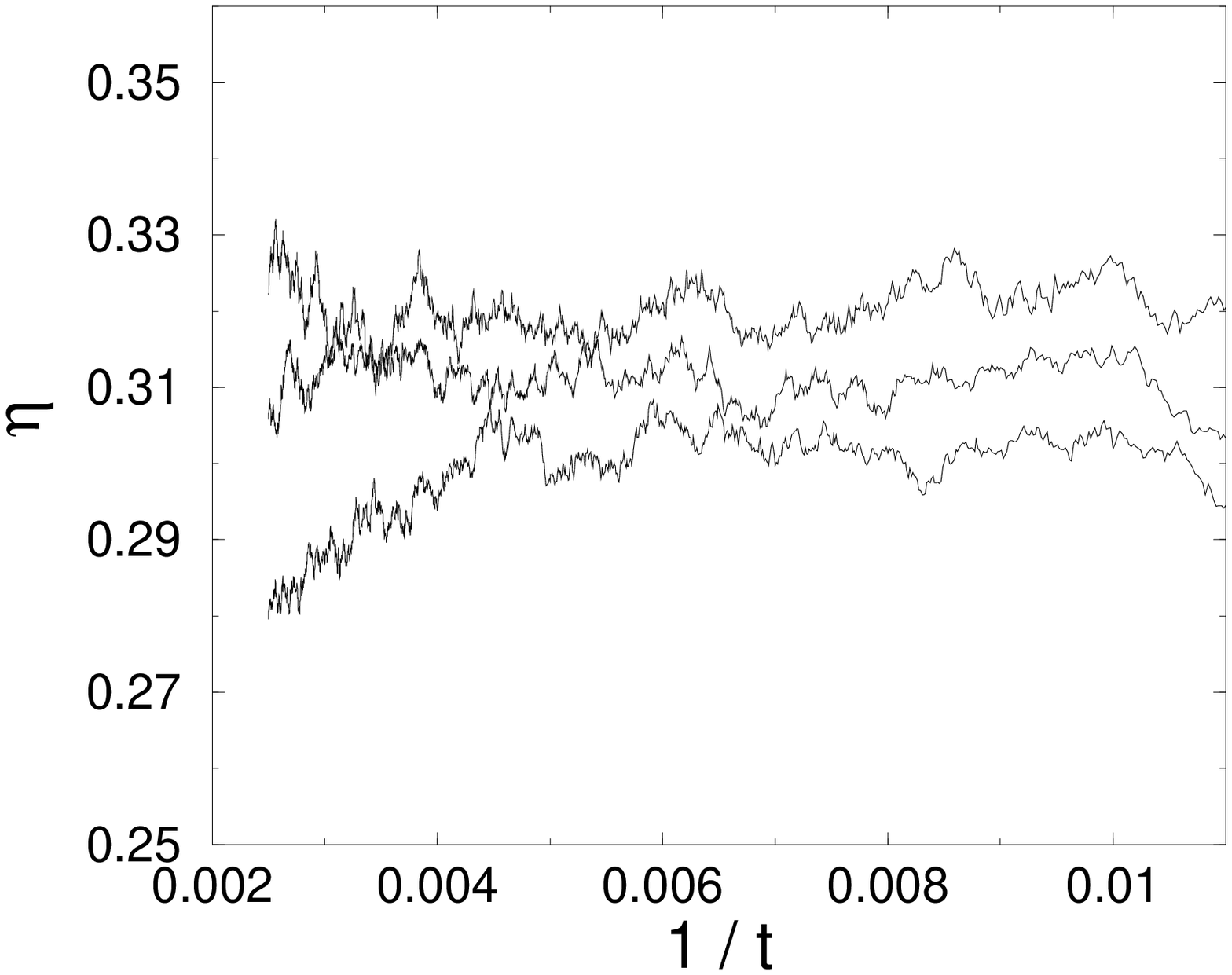,width=8.2cm,height=5.8cm}

\psfig{figure=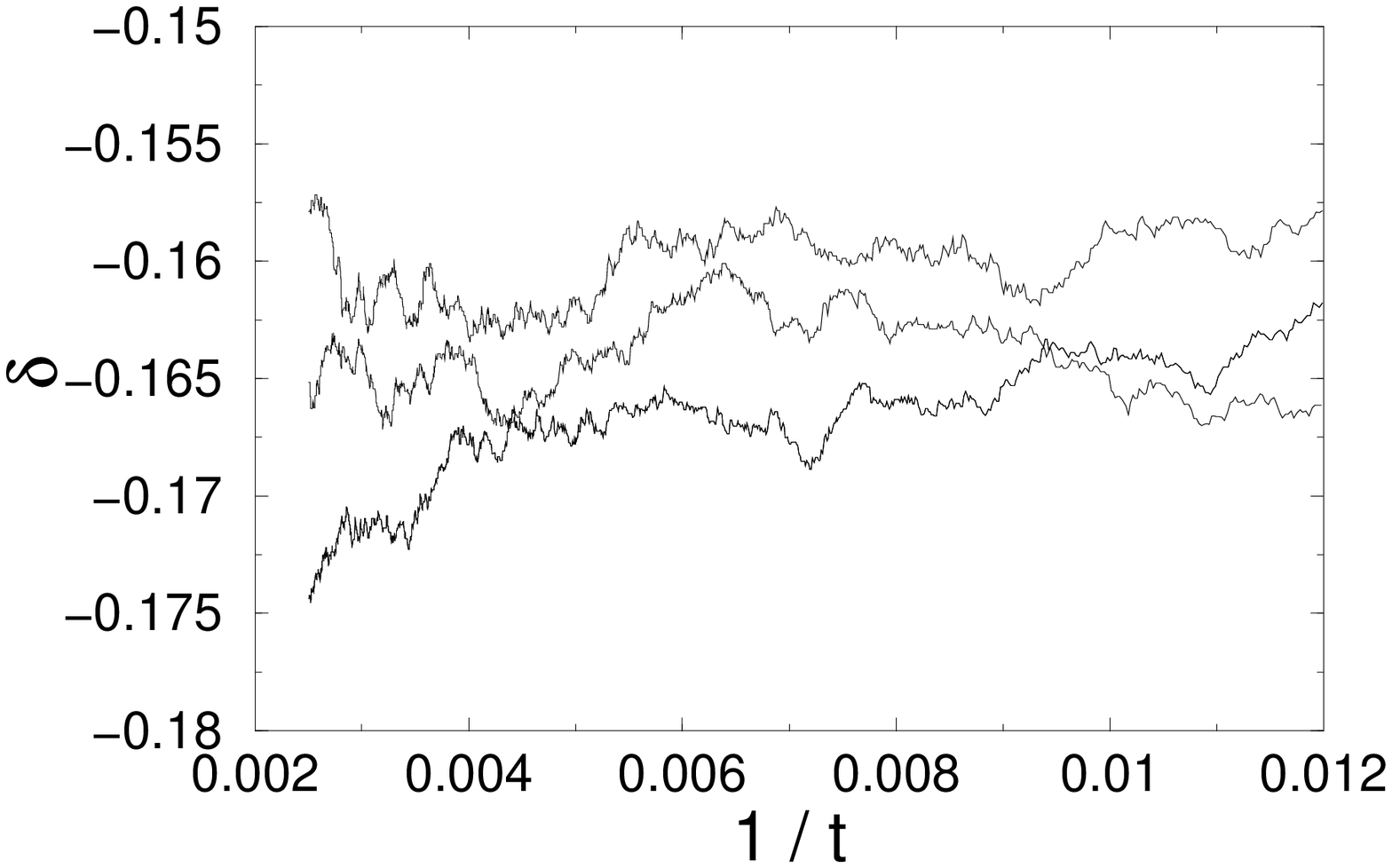,width=8.6cm,height=5.8cm}

\psfig{figure=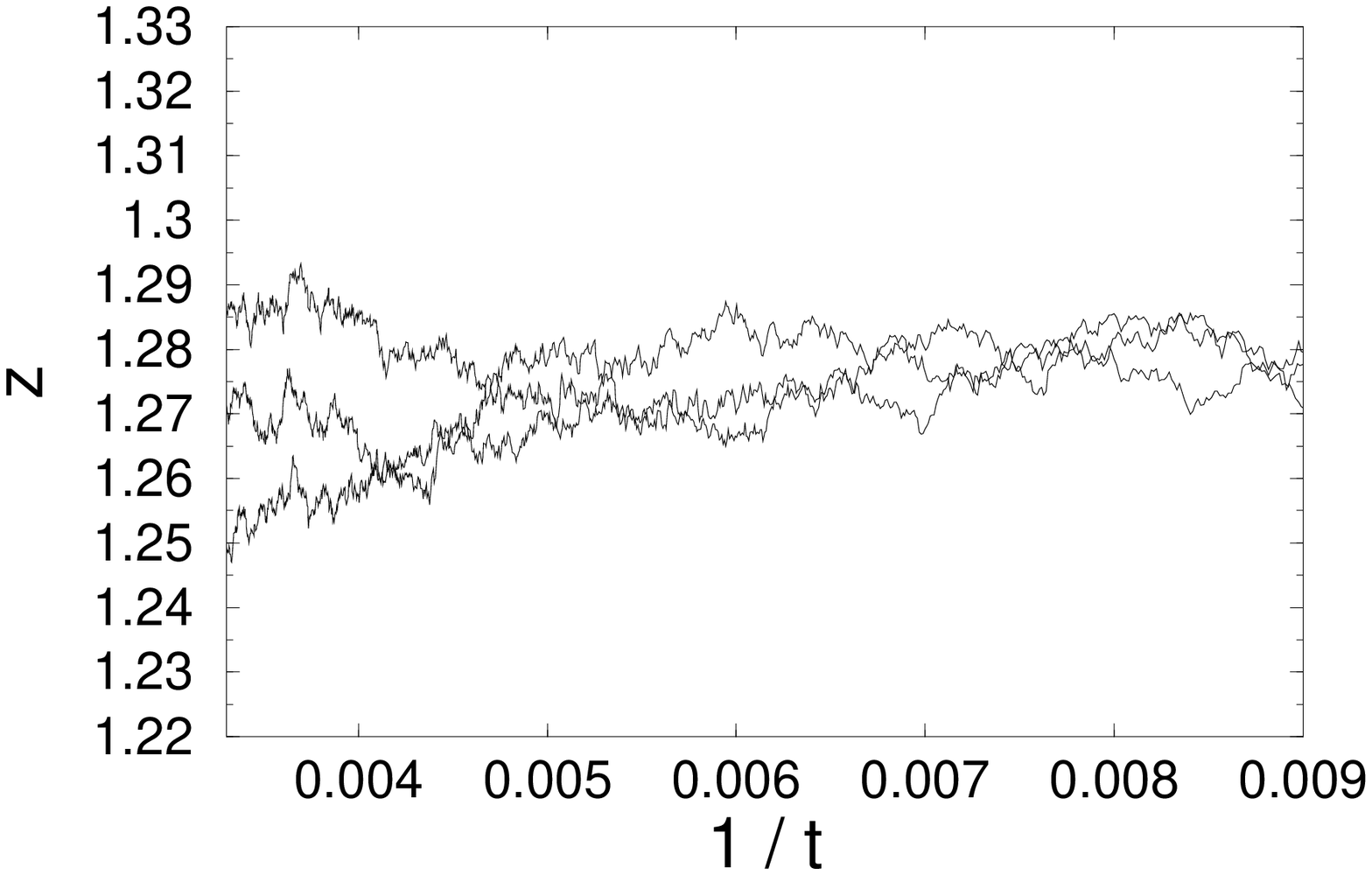,width=8.6cm,height=5.8cm}
\caption{ Dynamical critical exponents as functions of $1/t$. Different curves in each
panel correspond to $\tau = 3.06, 3.07, 3.08$ respectively from top to bottom.     
 }
\end{figure}

Figure (3) shows the results of the dynamical simulations.    
In different panels local slopes as defined in eq.(16) for different dynamical critical 
exponents have been depicted against $1/t$. The best estimation for the critical age based
on these graphs is $\tau_{c} \simeq 3.07(1)$. For the critical exponents I find
$\eta = 0.304(1)$, $\delta = 0.1653(1)$ and  $z = 1.272(1)$. These 
critical exponents are in good agreement with those of Directed Percolation and thus
the system belongs to the DP class.

\section {Conclusion}

In summary, a Non-Markovian version of the Contact Process is introduced. 
Particles have an age which determines when they are born and when they will be 
annihilated. Interesting oscillatory behaviors are observed in density
of particles and the same oscillations are also present in the Mean-Filed approximation.
Because of  the non-Markovian property of the model there is a time delay in the
mean-filed first-order differential equation which allows it to demonstrate 
oscillatory behaviors. 

Applying time-dependent Monte Carlo technique, critical properties of the absorbing 
phase transition has been investigated and shown to belong to DP universality class.
DP class is so extended to contain a non-Markovian model.

\acknowledgments

I would like to thank M. R. Ejtehadi and H. Hinrichsen for critical reading of the 
manuscript and stimulating comments and Stephan Haas for valuable  
supports. This work was supported by the National
Science Foundation Grant No. DMR-0089882.

\bibliographystyle{unsrt}

\end{document}